\documentclass[prb,aps]{revtex4}
\usepackage{amsmath, amssymb, color, graphicx, subfigure}
\newcommand{\mc}{\mathcal}
\begin{document}

\title{Photon dressed electronic states in topological insulators:
Tunneling and conductance}

\author{Andrii Iurov$^{1}$ \footnote[1]{E-mail contact: \href{aiurov@hunter.cuny.edu}{aiurov@hunter.cuny.edu}},
Godfrey Gumbs$^{1,2}$,  Oleksiy Roslyak$^{1}$, Danhong Huang$^{3}$}
\affiliation{$^1$ Department of Physics and Astronomy,
Hunter College, City University of New York 695 Park Avenue,
New York, NY 10065, USA  \\
$^{2}$ Donostia International Physics Center (DIPC), P de
Manuel Lardizabal, 4, 20018 San Sebastian, Basque Country, Spain \\
$^{3}$ Air Force Research Laboratory (ARFL/RVSS), Kirtland Air Force Base, NM 87117, USA}

\date{\today}

\begin{abstract}
The surface bound electronic states of three-dimensional topological
insulators, as well as the edge states in two-dimensional topological
insulators, are investigated in the presence of   a circularly polarized
light. The strong coupling between electrons and photons is found t
o give rise to an energy gap  as well as a unique energy dispersion of the
dressed states, different from both graphene and conventional two-dimensional
electron gas (2DEG). The effects of electron-photon  interaction,
barrier height and   width on the electron tunneling through a $p-n$
junction and on the ballistic conductance in topological insulators
are demonstrated by numerical calculations. A critical energy for an
incident electron   to  tunnel  perfectly through a barrier is predicted,
where electrons behave as either massless Dirac-like or massive
Schr\"odinger-like particles above or below this threshold value.
Additionally, these effects are compared with those in zigzag graphene
nanoribbons and a 2DEG. Both the similarities and the differences are demonstrated
and explained.
\end{abstract}
\maketitle

\section{Introduction}
\label{intro}

The unusual energy band structure of  topological insulators (TI),
as a novel class of quantum spin materials, has received  a
considerable amount of  theoretical attention in the last few
years\,\cite{HasanTI}. The energy dispersion is characterized
by an insulating gap in the three-dimensional (3D) bulk states
as well as by topologically protected conducting states localized
either around the two-dimensional (2D) surface for 3DTIs or
around the edge for 2DTIs\,\cite{Qi}. In this paper, we adopt
the conventional classification (2D/3D) for TIs based on their
geometry.

\medskip

Quantum spin Hall (QSH) topological states were discovered in
 HgTe/CdTe quantum wells (QWs). The  existence of these QSH states
 is determined by the QW thickness greater than  a critical value.
For films thicker than $6$\,nm, such QWs are exemplary 2DTIs but
become conventional insulators otherwise\,\cite{Ber}. Typical examples
of 3DTI include half-space Bi$_{1-x}$Sb$_x$ alloys as well as
Bi$_2$Se$_3$, Bi$_2$Te$_3$ and Sb$_2$Te$_3$ binary crystals.
The surfaces of these 3DTIs support spin-polarized Dirac cones
analogous to graphene\,\cite{Novoselov-main, geim}.
\medskip

\medskip

It has been shown that topological states may acquire an energy gap\,\cite{main_model,Finite_Size}. Since it is usually produced as
a geometrical gap, it requires a finite size along a given
direction. The energy gap depends on either the ribbon width
for a QW in 2DTIs or the separation between two surfaces of 3DTIs.
For example, a ribbon width of $200$\,nm produces a gap which
is about $0.5$\,meV. In this paper, we propose another approach
for generating an energy gap by coupling the topological surface
states to circularly polarized light. This type of dynamic gap
was predicted in graphene based on both a semiclassical
approach\,\cite{aoki_oki} and a quantum mechanical
formalism\,\cite{Kibis, Kibis2, K3}.

\medskip
The creation of an energy gap may lead to  a metal-insulator
phase transition. The conical dispersion of metallic graphene
has provided unimpeded electron tunneling through a $p$-$n$
junctions (Klein paradox)\,\cite{Novoselov-main, geim}. On the
other hand,  Klein-like total reflection has been predicted for
bilayer metallic graphene with its massive but still chiral electrons\,\cite{Katsnelson}. Chirality is shown to be the key
property for  total reflection. However,  perfect tunneling may
still be expected for certain values of the longitudinal momentum
of electrons in the barrier region because these transmission
resonances are not affected by the chirality.

\medskip
The light-induced energy gap is able to break  the chirality\,\cite{mine}
and suppresses the Klein effect in graphene. In this paper, we would
expect that a similar effect may occur in TIs because the helicity
of the topological states is also broken by an energy gap.
Our numerical results demonstrate a cross-over behavior from
Klein-like tunneling in a TI to  tunneling of conventional
a two-dimensional electron gas   (2DEG). Here, by Klein-like
we mean that the energy dispersion of topological states deviates
from the Dirac cone. For instance, in a 3DTI, there exists an
inherent mass term in the effective surface Hamiltonian,
which affects the Klein effect in TIs. The interplay between
the induced and inherent mass terms, as well as their competing
effects on the electron transmission, are the main subjects of
our investigation. In Ref.\,[\onlinecite{magntr}], tunneling and
transport problems in the presence of tilted uniform
magnetic and electric fields were studied, and we will
briefly discuss the effect of the dynamic gap for these cases.

\medskip
The rest of our presentation is organized as follows.
In Sec.\,\ref{Kibis}, we first discuss the dressed topological
states and obtain their energy dispersion and wave functions
for surface states and present an effective surface Hamiltonian
as well. Similar to graphene, the quantum field formalism predicts
a dynamic gap due to electron-photon coupling. The distorted
valence-band dispersion of the TI is calculated and compared to
that for  graphene with the main focus on broken chiral/helical
symmetry. In Secs.\,\ref{Transm} and \ref{2D}, we explore the
effect of an induced gap on electron transmission through a barrier
in a TI for incident energies either less or greater than  the barrier
height. Specifically, Sec.\,\ref{Transm} is devoted to tunneling
in 3DTIs along with comparisons to graphene and 2DEG, whereas
Sec.\,\ref{2D} deals with electron transmission in 2DTIs to complement
the results on electron tunneling in zigzag graphene nanoribbons (ZNRs).

\section{electron-photon interaction and dressed states}
\label{Kibis}

In this section, by including electron-photon coupling, we derive
 an effective Hamiltonian for surface states of TIs based on
 quantum field theory. Both the single-mode and double-mode
 optical fields are considered and their energy dispersions
  for dressed electron  states are  compared. Analogous
with graphene-like massless particles, the effect of massive
particles in TIs on electron states and tunneling are studies.
\medskip

Let us now consider electron-photon interaction on the surface
of a 3DTI. We obtain the dressed electronic states analytically
and investigate the tunneling properties of these states.
We first assume that the surface of the 3DTI is irradiated by
circularly polarized light with its quantized vector potential
given by

\begin{equation}
\label{vpsurf0}
\hat{\mathbf{A}}=\mc{F}_0\left(\mathbf{e}_+\,\hat{a}
+\mathbf{e}_-\,\hat{a}^{\dag}\right)\ ,
\end{equation}
where the left and right circular polarization unit vectors
are denoted by $\mathbf{e}_\pm=(\mathbf{e}_x
\pm i\,\mathbf{e}_y)/\sqrt{2}$,
and $\mathbf{e}_x$ ($\mathbf{e}_y$) is the unit vector in the $x$
($y$) direction. The amplitude of the circularly polarized light is
related to the photon angular frequency $\omega_0$ by
$\mc{F}_0\sim\sqrt{1/\omega_0}$. Here, we consider a weak field
 (energy $\sim \mc{F}^2_0$)  compared to the photon energy
 $\hbar \omega_0$. Additionally, the total number $N_0$ of
 photons is fixed for the optical mode represented by
 Eq.\,(\ref{vpsurf0}), corresponding to the case with  focused
 light incident on a portion of an optical lattice modeled by
 Floquet theory\,\cite{aoki_oki}.
\medskip

\begin{figure}[ht!]
\centering
\includegraphics[width=0.75\textwidth]{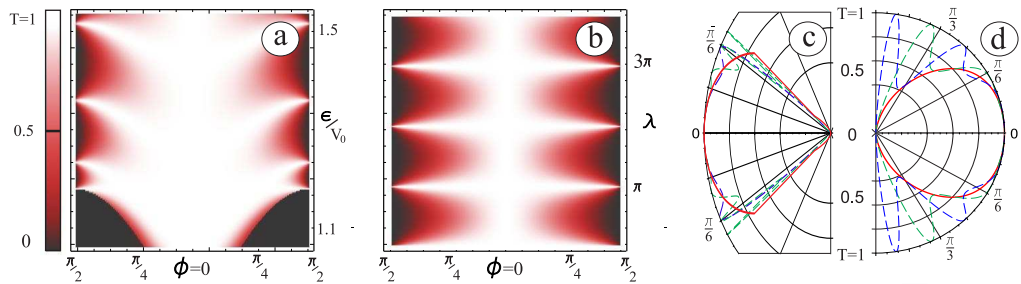}
\caption{(Color online) Transmission amplitude $\mc{T}$ for
Dirac-like surface states for a TI with no quadratic
term in the energy dispersion, $\mc{D}=0$. Panels
$(a)$ and (b) are  density plots of $\mc{T}(\phi,\,\varepsilon)$
for a potential $V(x)=V_0$ and width $W=50$\,nm as well as of $\mc{T}(\phi,\,\lambda)$ for  $V(x)=\lambda\,\delta(x)$ when
 $\varepsilon/V_0 = 1.5$, respectively. Panels $(c)$ and $(d)$
 show the $\phi$ dependence of $\mc{T}(\phi,\,\varepsilon)$
 with $W=50$\,nm at $\varepsilon/V_0 = 1.5$ and $\varepsilon/V_0 = 5$,
 respectively, where the red solid curves are for $W=50$\,nm,
 the blue dashed curves for $W=150$\,nm, and the green dashed
 curves for $W=250$\,nm.}
\label{grtun}
\end{figure}

The non-interacting Hamiltonian of 3DTI was derived in Ref.
\, [\onlinecite{DSTransp}] and we write it as

\begin{equation}
\mc{H}^{\rm esm}_{3D} =\mc{D}k^2\,\tensor{\mathbb{I}}_{[2]}
+\mc{A}\,\tensor{\sigma}
\cdot{\bf k} = \left({
\begin{array}{cc}     \mc{D} k^2  &         \mc{A}k_-  \\                                                                                   \mc{A}k_+   &       \mc{D}k^2 \end{array} }\right) \ ,
\label{esmHam}
\end{equation}
where $\tensor{\mathbb{I}}_{[2]}$ is a $2\times 2$ unit matrix,
$\tensor{\sigma}$ stands for the usual Pauli matrices,
${\bf k}=(k_x,\,k_y)$ is the in-plane surface wave vector
with respect to the $\Gamma$-point and $k_{\pm}=k_x \pm i\,k_y$.
For the 3DTI considered here, the group velocity is of the same
order of magnitude as graphene, i.e.,
$\mc{A}\sim\hbar v_F\sim 10^{-29}$\,J$\cdot$m.
It was shown that the leading quadratic term in Eq.\,(\ref{esmHam})
is necessary although the higher order terms with respect to
${\cal O}(k^2)$ may be neglected. The massless form of the Hamiltonian
in Eq.\,(\ref{esmHam}) with no quadratic term, $\mc{D} =0$,
formally coincides with graphene Dirac cones and retains all graphene
electronic properties. Specifically, for electron tunneling, these
properties include the absence of back-scattering for head-on collisions
(Klein paradox) as well as distinct tunneling resonances in the electron
energy distribution.

\medskip

The energy dispersion relation associated with Eq.\,\eqref{esmHam}
is $\varepsilon^{\rm surf}_{3D}=\mc{D}k^2+\beta \mc{A}\,\vert k \vert$,
where  $\beta = \pm 1$ is analogous with  \textit{pseudo-spin} in graphene.
Both $\mc{A}$ and $\mc{D}$ are independent of wave vector $k$.
This dispersion relation shows that the particle-hole symmetry is
broken by virtue of the massive $\mc{D}$-term.  For completeness,
the transmission amplitude of the massless topological states
(with $\mc{D}=0$) is presented in Fig.\,\ref{grtun}.
Comparing Figs.\,\ref{grtun}(a) and (b), we clearly see a significant
difference although the thickness of a potential barrier  is only $50$\,nm.
The effect of coupled dressed states on the tunneling is much
stronger for  a $\delta$-function barrier. At the same time,
the tunneling resonant peaks are broadened significantly compared
with graphene. From Figs.\,\ref{grtun}(c) and (d), we also find that
the angular distribution of transmission side-peaks at larger angles
displays a non-monotonic dependence on the barrier width $W$
for the higher scaled electron energy $\varepsilon/V_0=5$. In addition,
the broadening of resonant peaks at small angles is also
significant in comparison with graphene.
\medskip

The interaction with the optical mode may be introduced into the
Hamiltonian in Eq.\,\eqref{esmHam} via a standard transformation of
$\mathbf{k}\to \mathbf{k}+e\hat{\mathbf{A}}/\hbar$.
In Appendix\ A, we have shown that this transformation leads to the
following effective Hamiltonian, after the field correction has
been neglected,

\begin{equation}
\hat{\mc{H}}=\hbar\omega_0\,\hat{a}^{\dag}
\hat{a}+\mc{D}k^2\,\tensor{\mathbb{I}}_{[2]}
+2\zeta\mc{D}\left(k_+\hat{a}+k_-\hat{a}^{\dag}\right)
\tensor{\mathbb{I}}_{[2]}+\mc{A}\,\tensor{\sigma}\cdot{\bf k}
+2\zeta\mc{A}\left(\tensor{\sigma}_+\hat{a}
+\tensor{\sigma}_-\hat{a}^{\dag}\right)\ ,
\label{hamil1}
\end{equation}
where we have introduced a small parameter $\zeta=e\mc{F}_0/(\sqrt{2}\hbar)$
to describe the light-matter interaction. We  assume that the
optical mode accommodates a large number $N_0$ of photons with
$N_0 \gg 1$. Consequently, all the terms ofder $\zeta^2\sim{\cal O}(1/N_0)$
may be neglected. Under these conditions, the energy dispersion
associated with Eq.\,(\ref{hamil1}) becomes

\begin{equation}
\varepsilon^{\rm surf}_{3D}(\Delta) = N_0\,\hbar \omega_0
+ \mc{D} k^2 + \beta \sqrt{\Delta^2
 + \left({\mc{A}k}\right)^2}
\label{singlemodedispersion}
\end{equation}
with $\beta=\pm 1$ and the induced energy gap defined by

\begin{equation}
\Delta = \sqrt{\mc{W}_0^2+(\hbar \omega_0)^2} - \hbar \omega_0 \sim \hbar \omega_0\left(\frac{\alpha^2}{2}\right)\ ,
\end{equation}
where $\alpha=\mc{W}_0/(\hbar\omega_0)$ and $\mc{W}_0$ is the
electron-photon interaction energy. For the upper subband with $\beta=1$
in Eq.\,\eqref{singlemodedispersion}, the energy gap is related
to the effective mass around ${\bf k}=0$ through
$2 m^\ast_\Delta=\hbar^2/\left[{\mc{A}^2/ (2 \Delta) + \mc{D}}\right]$,
where the photon dressing decreases the effective mass. This is in
contrast with single-layer graphene, where electron-photon interaction
leads to an  an effective mass. A similar phenomenon
 on the  effective mass reduction is also found in bilayer
 graphene under the influence of  circularly polarized light.
The dressed state wave function corresponding to
Eq.\,(\ref{singlemodedispersion}) is given by

\begin{gather}
\Phi^{\bf k}_{\rm e-ph}(x,\,y)=\frac{1}{\sqrt{1+\gamma^2(\beta)}}
\left({ \begin{array}{c}    1 \\
\gamma(\beta)\texttt{e}^{i \phi}  \end{array}  }\right)
\texttt{e}^{ik_x x + ik_y y}\ ,
\end{gather}
where $\gamma(\beta)=\mc{A}k/[\Delta+\beta
\sqrt{\Delta^2+(\mc{A}k)^2}]$ and $\phi=\tan^{-1}(k_y/k_x)$.
The energy dispersions associated with the
Hamiltonian in Eq.\,\eqref{hamil1} [also see Eq.\,
\eqref{4ham}] for two-mode light interaction with electrons are given by

\begin{gather}
\label{abb}
\varepsilon_{\{N_0 \uparrow, N_0 \downarrow,\,N_0
+1 \uparrow, N_0+1 \downarrow\}}(k,\,\Delta) =\left({N_0
+\frac{1}{2}}\right) \hbar \omega_0 + \mc{D} k^2
\pm \sqrt{\mathbb{C}_1(k,\,\Delta) \pm \sqrt{\mathbb{C}_2(k,\,\Delta)}} \ ,
\end{gather}
where a doubled state space is used for spanning the Hamiltonian,

\begin{gather}
\mathbb{C}_1(k,\,\Delta)=\left({\hbar \omega_0 / 2}\right)^2
+\zeta \nu \mc{A} \mc{D} k^2 + [\zeta^2 \mc{D}^2 + \mc{A}^2(1 + 5/2
\nu^2 )] k^2 \ ,\\
\notag
\mathbb{C}_2(k,\,\Delta)=\zeta^2 \mc{A}^2 \mc{D}^2 k^4
(1+ \nu^2) + 4 \nu (\zeta \nu \mc{D}+\mc{A}) \mc{A}^3 k^4\\
 - 3 \nu (2 \zeta \mc{D}+\nu \mc{A}) \mc{A} \hbar\omega_0 \Delta k^2 + 4(\hbar\omega_0)^2\,[\Delta^2 + (\mc{A}k)^2] \ ,
\end{gather}
and $\nu\sim\alpha/2$.

\begin{figure}[ht!]
\centering
\includegraphics[width=0.75\textwidth]{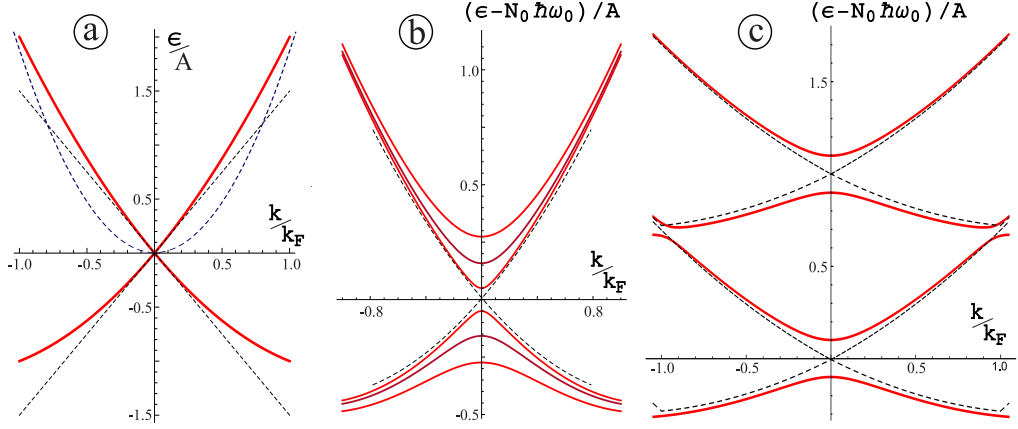}
\caption{(Color online) Energy dispersion (red solid curves)
for the effective surface model of 3DTI: (a) without light-electron
interaction [$\alpha=0$, $\mc{D}k^2_F/(\hbar\omega_0)
=\mc{A}k_F/(\hbar\omega_0)= 0.2$]; (b) for single-mode dressed states [$\alpha
= 0.05$ (inner), $0.07$ (middle), $0.1$ (outer)]
in Eq.\,\eqref{singlemodedispersion}; and (c) for two-mode dressed states in Eq.\,\eqref{abb}.
Here, the black dashed  lines in (a) represent the Dirac cones
and a parabola, while
the black dashed curves in (b) and (c) indicate the asymptotical behaviors for $\alpha=0$.}
\label{kd}
\end{figure}

As $\zeta \mc{D} \to 0$ and  $\nu k\to 0$, the two-mode dressed states
become decoupled and are simply given by

\begin{gather}
\varepsilon_{\{N_0 \uparrow,\,N_0 \downarrow\}}(k,
\,\Delta) = N_0\,\hbar\,\omega_0 + \mc{D} k^2 \pm
\sqrt{\Delta^2+(\mc{A}k)^2} \ ,\\
\varepsilon_{\{N_0+1 \uparrow,\,N_0+1 \downarrow\}}(k,\,
\Delta) = (N_0+1)\,\hbar \omega_0 + \mc{D} k^2 \pm
\sqrt{\Delta^2+(\mc{A}k)^2} \ .
\end{gather}
On the other hand, for single mode dressed states, the effect
due to the electron-photon interaction is quite similar to graphene,
except that the energy gap varies as  $\alpha^2$. However, this
dependence becomes negligible under low -intensity light
illumination. The energy dispersion relations for single  and
double-mode dressed states of 3DTIs are presented in Fig.\,\ref{kd}.
Comparing Figs.\,\ref{kd}(a) and (b), we find that an energy gap
is opened at ${\bf k}=0$ due to photon dressing, and the Dirac cone
is well maintained except for large $k$ values.  In contrast,
for double-mode dressed states in Fig.\,\ref{kd}(c),  additional
mini-gaps appear at the Fermi edge and new saddle points are
formed at ${\bf k}=0$ due to strong coupling between dressed states
with different pseudo-spins. These new mini-gaps and the saddle
points prove to have a significant effect on electron
tunneling.
\medskip

The full expression for the 3DTI Hamiltonian which includes
the $z$ dependence (perpendicular to the surfaces) may also
be related to dressing and can be expressed as\,\cite{main_model}
(see Eq. (\ref{ap2}in the Appendix)

\begin{gather}
\label{3dH}
\hat{\mc{H}}_{3D}({\bf k}_\bot,\,z)=
\hat{\mc{H}}^{(1)}_{3D}(z)+
\hat{\mc{H}}^{(2)}_{3D}({\bf k}_\bot) \ ,
\end{gather}
where

\begin{gather}
\hat{\mc{H}}^{(1)}_{3D}(z) =\left({\mc{C}-
\mc{D}_z\partial_z^2}\right)
\tensor{\mathbb{I}}_{[4]}+\left({ \begin{array}{cc}    (\mc{M}+\mc{B}_z\partial_z^2)\, \tensor{\sigma}_3 - i \mc{A}_{z}\partial_z\,\tensor{\sigma}_1  &  0 \\
 0  &  (\mc{M}+\mc{B}_z\partial_z^2)\,\tensor{\sigma}_3 + i \mc{A}_{z}\partial_z\,\tensor{\sigma}_1
\end{array} }\right) \ , \\
\hat{\mc{H}}^{(2)}_{3D}({\bf k}_\bot) = -\mc{D}_{\bot} k^2
\tensor{\mathbb{I}}_{[4]} -  \mc{B}_{\bot} k^2\,\tensor{\sigma}_3 \otimes\tensor{\mathbb{I}}_{[2]}   +  \left({
\begin{array}{cc}                     0                &            \mc{A}_{\bot}k_-\,\tensor{\sigma}_1 \\
\mc{A}_{\bot}k_+\, \tensor{\sigma}_1                &            0
\end{array} }\right) \ .
\label{part2}
\end{gather}
In this notation, ${\bf k}_\bot=(k_x,\,k_y)$,
$\mc{C}$, $\mc{M}$, $\mc{A}_z$ , $\mc{B}_z$, $\mc{D}_z$
are parameters in the Kane ${\bf k}\cdot{\bf p}$ model for
bulk states, and $\mc{A}_\bot$ , $\mc{B}_\bot$, $\mc{D}_\bot$
are the parameters for surface states. The corresponding energy
dispersion relations are linear and gapless for semi-infinite
samples (half-space). However, for a TIs of finite-width,
an energy gap is opened due to the finite-size effect.
We further find that, similar to graphene,  electron-photon
interaction may modify the  energy gap due to additional
contributions from dressing.

\section{Tunneling and Ballistic Transport in 3DTI}
\label{Transm}

In this section, we compare surface states of a 3DTI with
bilayer graphene in order to find out how w the
quadratic term in their energy dispersions influences
tunneling. The electron tunneling behaves as massless Dirac
 fermions for high incident energies above a critical value,
 which has been compared with electrons in a graphene layer.
On the other hand, the electron tunneling behaves as Schr\"odinger
massive particles at lower incident energies below this critical
value, which is similar to the 2DEG. Additionally, the modification
from a dressed state energy gap to electron tunneling is investigated
 for both low and high incident energies.
\medskip

Let us first turn to the tunneling problem associated with the
Hamiltonian in Eq.\,\eqref{esmHam} in the absence of electron-photon
coupling. The main input of the $p$-$n$ junction tunneling is
the electron wave function\,\cite{Katsnelson,castroneto}

\begin{equation}
\Psi_{\bf k}(x,\,y)= \frac{1}{\sqrt{2}} \left({
\begin{array}{c}            1 \\
      \beta  \texttt{e}^{i \phi}
      \end{array}  }\right) \texttt{e}^{ik_x x+ik_y y} \ ,
\end{equation}
where $\beta=\pm 1$, as before,  and $\phi=\tan^{-1}(k_y/k_x)$.
This wave function is simple and chiral, and an eigenstate
of the projection of the electron momentum operator along the
pseudo-spin direction (chirality/helicity operator) given by
$\hat{h}=\tensor{\sigma}\cdot\mathbf{p}/(2p)$.
For the tunneling process along the $x$ direction, the
transverse momentum $\hbar k_y$ is conserved, whereas the
longitudinal wave vector component $k_{x,2}$ in the barrier
region with $V(x)=V_0$ is determined by

\begin{gather}
\label{kx2gen}
k_{x,2}^2 = \left({\frac{-\beta \mc{A} + \beta\sqrt{ \mc{A}^2+ 4\mc{D}(\varepsilon-V_0)}}{2 \mc{D}} }\right)^2 -k_y^2  \ .
\end{gather}
Equation\ \eqref{kx2gen} yields not only propagating but
also evanescent modes in the barrier region. Moreover,
because of the quadratic energy dispersion, both the wave
functions and their derivatives need to be matched at boundaries,
similar to the tunneling problem with respect to bilayer
graphene\,\cite{Katsnelson}.

Electron tunneling and Andreev reflection for the     full
3DTI Hamiltonian was studied in Ref.\,\onlinecite{andreevr}.
In the present case, we have $\mc{A}/\mc{D} \gg k$. Consequently,
 we may neglect  the evanescent contributions. By neglecting the
 higher-order terms, Eq.\,\eqref{kx2gen} is simplified to

\begin{equation}
\label{kx2}
k_{x,2}=\sqrt{\frac{(\varepsilon-V_0)^2}{2
\mc{D}(\varepsilon-V_0)+\mc{A}^2}-k_y^2}\ .
\end{equation}
The result for the special case of massless Dirac fermions
 of graphene may be directly obtained from the above equation
 after setting $\mc{D} \to 0$.

\medskip

Clearly, from Eq.\,\eqref{kx2}, the electron transmission
varies with the incoming particle energy  as well as the  angle of
incidence. There exists a \textit{critical energy}
$\varepsilon_{\rm cr} \ll V_0$ above  which the transmission
behaves like Dirac electrons in graphene. However, particles with
incoming energies below this critical value are transmitted like
normal Schr\"{o}dinger electrons in 2DEG. For  head-on
collisions with $k_y=0$, the critical energy is calculated to be

\begin{equation}
\varepsilon_{\rm cr} = V_0- \frac{\mc{A}^2}{2\mc{D}} \ .
\end{equation}

\medskip

The dimensionless two-terminal tunneling conductance,
$g(\varepsilon)$, may be calculated from\,\cite{Barb_review}

\begin{equation}
\label{tuncondg}
g(\varepsilon)=\frac{\mc{G}(\varepsilon)}{2 \mc{G}_0} = \frac{1}{2} \int\limits_{-\pi/2}^{\pi/2}\,{\cal T}(\phi,\,\varepsilon)\,\cos(\phi)\,d\phi\ .
\end{equation}
The physical meaning of Eq.~\eqref{tuncondg} is the decrease of the
electron conductance in the presence of the barrier due to ${\cal T}(\phi,\,\varepsilon)\leq 1$, where ${\cal T}(\phi,\,\varepsilon)$
is the transmission probability. According to Ref.\,
[\onlinecite{Tw}], $\mc{G}_0$ in the case of $\mc{D}=0$ may
 be estimated using

\begin{equation}
\mc{G}_0=\left(\frac{e^2}{h}\right)\,\frac{L_y}{2 \pi}\,\int\limits_{-k_F}^{k_F}\frac{d k_y}{\cosh^2(k_y L_x/2)} \ ,
\end{equation}
where $k_F$ is the Fermi wave number and $L_x$ and $L_y$ are the
normalization length and  width of the sample. When $k_F L_x \ll 1$,
we have $\tanh(k_F L_x/2)\approx k_F L_x/2$ and   obtain in a
straightforward way  that
$\mc{G}_0 =  (2 e^2/ h)\,(L_y/2 \pi)\,(\varepsilon_F /\mc{A})$
for the Dirac cone with $\varepsilon_F=\mc{A}k_F$.
In the presence of a small energy gap $\Delta \ll \mc{A} k_F$,
on the other hand, $\mathcal{G}_0$ is modified to

\begin{equation}
\mc{G}_0 =  \left(\frac{2 e^2}{h}\right)\, \frac{L_y \sqrt{\varepsilon_F^2-\Delta^2}}{2 \pi \mc{A}}\ ,
\end{equation}
where $\varepsilon_F$ is the Fermi energy at $k=k_F$.

\medskip

\begin{figure}[ht!]
\centering
\includegraphics[width=0.65\textwidth]{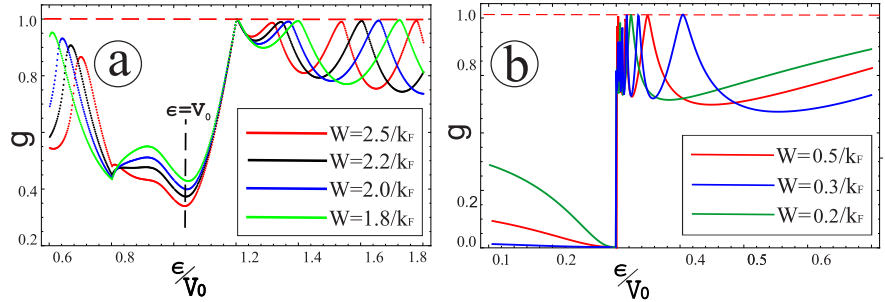}
\caption{(Color online) Two-terminal tunneling conductance
for the effective surface model of 3DTI with different barrier
widths, where $\mc{D}k_F/\mc{A}=0.4$.
In panels (a)  and (b), we present  numerical results for tunneling
conductance as functions of the incident particle energy for the
case of Dirac cone and 3DTI.}
\label{tunnel}
\end{figure}

The calculated results based on Eq.\,\eqref{tuncondg} are presented in Fig,\,\ref{tunnel}. Comparing  Figs.\,\ref{tunnel}(a) and (b),
 we find that the existence of the critical energy in (b) plays
 a crucial role in the electron tunneling. A series of resonant
 peaks occur above the critical energy $\varepsilon_{\rm cr}$
 in (b) and above the barrier height $V_0$ in (a) although
 these two values are quite different. The former is unique to
 surface states of a massive particle in 3DTI while the latter
 is related to the transition from Klein tunneling to a regular
 one for massless fermions in graphene. It is more interesting
 to notice that the electron tunneling below $V_0$ in (a) and
 below the critical energy in (b) is also qualitatively
 different, where electrons behave like Dirac fermions for
 Kelin tunneling in (a) and like Schr\"odinger particle
  for  evanescent wave tunneling in (b). Transition when $\varepsilon=\varepsilon_{\rm cr}$ is extremely sharp in (b).

\medskip

It is very helpful to compare the results obtained in this section
with bilayer graphene having quadratic dispersion.
For  bilayer graphene, its lowest energy states are described
by the Hamiltonian\,\cite{Katsnelson}

\begin{equation}
\hat{\mc{H}}_{\rm blg}=\frac{\hbar^2}{2 m_{b}}
\left({k_-^2\,\tensor{\sigma}_+ + k_+^2\,\tensor{\sigma}_- }\right)\ ,
\end{equation}
where $m_b$ is the effective mass of electrons in the barrier region.
In this case, the longitudinal wave vector component in the barrier
region is given by

\begin{equation}
k_{x,2}^{b}=\beta^\prime\,\sqrt{2m_b\beta (\varepsilon - V_0)
- k_y^2} \ ,
\end{equation}
where $\beta^\prime,\,\beta=\pm 1$. An evanescent wave having a decay
rate $\kappa_{b}$ can coexist with a propagating wave having  wave
vector $k_{x,2}$ such that $k_y^2+k_{x,2}^2=k_y^2-\kappa_{b}^2=2 m_b\beta (\varepsilon - V_0)$.
This implies that the evanescent modes should be taken into account
simultaneously. For bilayer graphene, both wave functions and thei
derivatives must be continuous at the interfaces. Consequently, the
Klein paradox persists in bilayer graphene for chiral but massive
particles. However, one finds  complete reflection, instead of
complete transmission, in this case. This effect has direct
links with the specific electron-hole conjugation, i.e.
$k_{x,2}\to i\kappa_b$, in the barrier region\,\cite{Katsnelson}.

\medskip

To understand the physics for electron tunneling in bilayer
graphene, we first present  analytical results for single
layer graphene, i.e. $\mc{D}=0$. By taking into account  all
four modes from Eq.\eqref{kx2gen}, for a $\delta$-potential
barrier, we obtain the transmission probability as\,\cite{Barb_review}

\begin{equation}
\mc{T}_{\delta,\,g}=\frac{1}{1 +\sin^2(\lambda)\,\tan^2(\phi)} \ ,
\label{deltag}
\end{equation}
 reproduces the Klein paradox for the head-on collision
corresponding to $\phi=0$. However, the periodic dependence
of the transmission on the scaled barrier strength $\lambda$
in Eq.\,\eqref{deltag} is non-trivial. Such perfect tunneling
with $\mc{T}_{\delta,\,g}=1$ also exists for a set of $\lambda$
values when $\sin(\lambda)=0$ is satisfied. Furthermore, this
prediction is  consistent with  finite-width barrier tunneling
[see Fig.\,\ref{grtun}\,(c)]. In fact, the expression in
Eq.\,\eqref{deltag} may be derived directly from the general
result for  electron transmission through a very high potential
barrier with $V_0 \gg \varepsilon$, that is,

\begin{equation}
\mc{T}_g=\frac{\cos^2(\phi)}{1-\cos^2(k_{x,2}W)\,
\sin^2(\phi)}=\frac{1}{1+\sin^2(k_{x,2}W)\,\tan^2(\phi)} \ ,
\label{2degs}
\end{equation}
where $k_{x,2} \sim -V_0/\mc{A}$ is used for $V_0
\gg \varepsilon$. It is clear that $\lambda$ in
Eq.\,\eqref{deltag} plays the role of $k_{x,2}W$ for a
finite barrier width $W$.

\medskip

For a conventional 2DEG with a $\delta$-function potential
barrier\,\cite{Flugge}, its transmission amplitude is given by
$\mc{T}_{\lambda}=2\hbar^2 \varepsilon/(2 \hbar^2
\varepsilon + m^\ast \lambda^2 W^2)$.
Here, we consider a head-on collision with
$k_x=\sqrt{2 m^\ast \varepsilon / \hbar^2}$ and
$m^\ast$ is the electron effective mass.
One may easily see that for $\lambda \to 0$
(or a very-thin barrier layer)  complete transmission
($\mc{T}_{\lambda}\to  1$) can be obtained. When
 $\lambda\to\infty$, on the other hand, one gets  complete
reflection.

\medskip

\begin{figure}[ht!]
\centering
\includegraphics[width=0.95\textwidth]{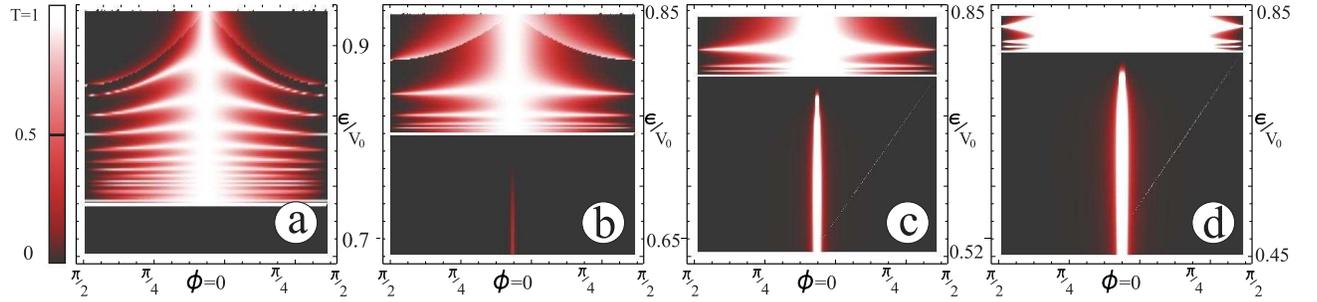}
\caption{(Color online) Density plots of the transmission
amplitude $\mc{T}(\phi,\,\varepsilon)$
for  bare electron states with energy gap  $\Delta=0$ in the effective
surface model for 3DTI with $(\mc{D}/\mc{A})\,k=0.4$. Panels (a),
(b), (c) and (d) correspond to $k_{1,\rm max}W= 2.5$, $1.1$,
$0.95$ and $0.75$.}
\label{tr}
\end{figure}

Figure\ \ref{tr} presents our numerical results for the
transmission amplitude $\mc{T}(\phi,\,\varepsilon)$
for several   values of barrier width in the absence of photon
dressing or when the coefficient $\Delta=0$ in the energy
dispersion relation. From this figure, we find that the
coupling between dressed states with different pseudo-spins
is very strong for a double-mode optical field, similar
in nature to the result in Fig.\,\ref{kd}(c). Here,  Dirac-like
tunneling above a critical energy may be seen for a thick
barrier layer, as described by Eq.\,\eqref{2degs}.
 Schr\"odinger-like tunneling below the critical energy, on
 the other hand, may only be observed for a relatively thin
 barrier layer under the normal-incidence condition, as discussed
 for 2DEG with $\lambda\to 0$. With decreased $W$,  electron
 tunneling below the critical energy is gradually enhanced
 as $\phi\to 0$. From direct comparison between Fig.\,\ref{tr}
 and Fig.\,\ref{grtun}(a) we know that the major effect of the
 massive $\mc{D}$-term on the electron tunneling is associated
 with the occurrence of a critical energy below which no
 significant electron tunneling is expected for a thick barrier
 layer. In addition, the critical energy shifts up with decreasing $W$ and the contribution from the evanescent mode to the transmission is found finite as long as the barrier width $W$ meets the condition $W\vert {\rm Im}(k_{x,2}) \vert<1$.
\medskip

\begin{figure}[ht!]
\centering
\includegraphics[width=0.95\textwidth]{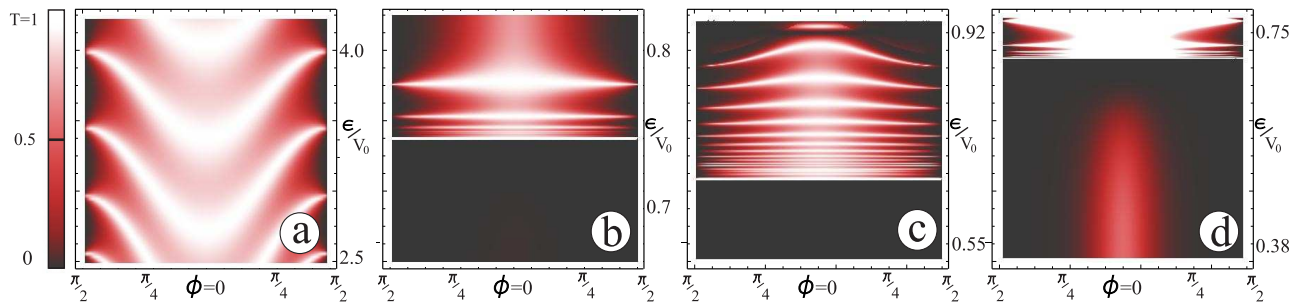}
\caption{(Color online) Density plots of the transmission amplitude $\mc{T}(\phi,\,\varepsilon)$ for the electron dressed states in the
effective surface model of 3DTI with $\Delta/V_0 = 0.2$, $\mc{D}=0$
(Dirac cone) for panel (a) and $(\mc{D}/\mc{A})\,k = 0.15$ for all
other panels. Panels (a), (b), (c) and (d) correspond respectively
to $ k_{1,\rm max}W = 1.5$, $1.3$, $1.1$ and $0.85$.}
\label{trk}
\end{figure}

 We now turn to the model for the 3DTI surface  with $\Delta\neq 0$.
 For this, the longitudinal momentum of electron dressed states
 may be approximated by

\begin{equation}
k_{x,2}^{\Delta}=\sqrt{\frac{(\varepsilon-V_0)^2
-\Delta^2}{2\mc{D}(\varepsilon-V_0)+\mc{A}^2}-k_y^2} \ .
\label{dres}
\end{equation}
Numerical results for the transmission amplitudes based on
Eq.\,\eqref{dres} are presented in  Fig.\,\ref{trk},
where the same parameters  were chosen as those in Fig.\,\ref{tr}.
Comparing Fig.\,\ref{trk}(a) with Fig.\,\ref{grtun}(a) for
$\mc{D}=0$,  we  find that the effect of a photon-induced energy gap
ia to produce  additional side peaks in the angle distribution of
$\mc{T}(\phi,\,\varepsilon)$. After a massive $\mc{D}$-term is introduced,
as presented in Figs.\,\ref{trk}(b)-(d), the side peaks are significantly
 suppressed and a critical energy appears. At the same time, the
 Schr\"odinger-like electron tunneling below this critical energy
 is also partially suppressed for a thin barrier layer even under
 the normal-incidence condition by comparing Fig.\,\ref{trk}(d) with Fig.\,\ref{tr}(d).

\section{Electron Tunneling in 2DTI and ZNR}
\label{2D}

In this section, we compare edge states of a 2DTI with a zigzag graphene
nanoribbon in order to find any similarities resulting from an induced
energy gap in their energy dispersions. At the same time, the dissimilarities
 between the 2DTI and an armchair graphene nanoribbon is also discussed.
 Additionally, the effect of the decay of electron wave functions away
 from the edges on the electron tunneling are investigated  both
 inside and outside the barrier region.

\medskip

According to the Kane model for a HgTe/CdTe semiconductor quantum
well\,\cite{Qi,Finite_Size}, the 2DTI system may be effectively
described by the following $4 \times 4$ matrix Hamiltonian, i.e.,

\begin{equation}
\label{H2Dmain}
\mc{H}_{2D}(k_x,\,y)= \left({ \begin{array}{cc} \tensor{h}_B(k_x,\,y) & 0 \\
0 & \tensor{h}^\dag_B(k_x,\,y) \end{array}  }\right)  \ ,
\end{equation}	
where

\begin{equation}
\label{H2Dmain2}																				 \tensor{h}_{B}(k_x,\,y)= \left({ \begin{array}{cc} \mc{C} + \mc{M} -(\mc{D}+\mc{B})(k_x^2-\partial_y^2)       &     \mc{A}(k_x - i\partial_y)  \\
\mc{A}(k_x + i\partial_y) &  \mc{C} - \mc{M} -(\mc{D}-\mc{B})(k_x^2-\partial_y^2) \end{array}  }\right)  \ ,
\end{equation}
where $k_x$ is assumed small. Additionally,  we assumed  translational
symmetry along the x-axis so that $k_x\to  -i \partial_x$.
The coefficients $\mc{A}$, $\mc{B}$, $\mc{C}$, $\mc{D}$ and $\mc{M}$
are expansion parameters within the Kane ${\bf k}\cdot{\bf p}$ model.
Their role in the energy dispersion is such that  $\mc{M}$ is usually
referred to as Dirac mass and $\mc{B}$ is called the Newtonian mass.
We note that $\mc{M}$ changes sign at the critical thinness
$d_{\rm cr} \simeq 6.1$\,nm. For a  layer with $d>d_c$, $\mc{M}$
is negative which is  referred to as an  \textit{inverted-type} TI.
\medskip

We note that the Hamiltonian  in Eq.\,\eqref{H2Dmain} is block-diagonal,
for which we expect two independent wave functions,
$\Psi_{+}$ and $\Psi_{-}$, corresponding to different directions
 of the electron spin projection. It is straightforward to show
  that $\Psi_{+}$ and $\Psi_{-}$ are related to each other
by a time-reversal operator\,\cite{Finite_Size}, i.e.
$\Psi_{-}=-i\tensor{\sigma}_{y} \Psi_{+}^\dag$. This implies that
any transmission associated with $\Psi_{+}$ must accompany
another transmission of $\Psi_{-}$ in the opposite direction
along the same edge of the 2DTI. For that reason, the eigenstates
$\Psi_{\pm}$ are usually referred to as helical edge states\,\cite{Qi}.
For all the calculations which follow, we made use of the fact
that $\mc{C}$ only appears along the diagonal of the matrix in
Eq.\eqref{H2Dmain2}. By making the replacement
$\varepsilon \to  \varepsilon - \mc{C}$, one  may show in a
straightforward way  that the $\mc{C}$-term is irrelevant to
electron transmission.
\medskip

Before solving the eigenvalue problem, we estimate\,\cite{Konig_exp}
all quantities and coefficients in the Hamiltonian in Eqs.\,\eqref{H2Dmain}
and \eqref{H2Dmain2} at $d=5$\,nm ($<d_{\rm cr}$). The range of
considered wave vectors is chosen as $k= 10^{-2}$\AA$^{-1}\ll k_F$.
In that range, the parameters appearing in the Hamiltonian
are:\,\cite{Qi} $\mc{C} \simeq 10^{-2}$\,eV, $\mc{A}k \simeq \mc{B}k^2\simeq\mc{D}k^2
\simeq 10^{-2} eV$. Consequently, each term in the Hamiltonian in
Eq.\,\eqref{H2Dmain2} has  the same order of magnitude and none
of them should be neglected. Since $\mc{B} \simeq \mc{D}$, we
get $\mc{B}_{+}\equiv\mc{B}+\mc{D} \simeq 2 \mc{B}$ and
$\mc{B}_{-}\equiv\mc{B} - \mc{D} \ll \mc{B}$.

\medskip

Apart from those considerations concerning the
quantities and coefficients in the Hamiltonian in Eq.\,\eqref{H2Dmain},
 the matrix may be separated into two parts. One
 part depends only on the $y-$coordinate
whereas the other    one is $k_x$-dependent. The eigenvalues are
determined from the following secular equation

\begin{equation}
\left({\mc{M} - \mc{B}_{+} (k_x^2 - \xi^2) -
\varepsilon}\right)\left({\mc{M} - \mc{B}_{-} (k_x^2
- \xi^2) - \varepsilon}\right)
=  - \mc{A}^2 (k_x^2 - \xi^2) \ ,
\label{mast}
\end{equation}
where $\xi$ characterizes the decay rate of edge state electron
wave functions with a finite width in the $y$ direction.
After neglecting the $\mc{B}_{-}$-term, we obtain from
Eq.\,\eqref{mast}

\begin{equation}
\xi^2 = k_x^2 - \frac{(\varepsilon - \mc{M})^2}
{\mc{A}^2-\mc{B}_{+} (\varepsilon - \mc{M})} \ .
\end{equation}
The electron wave function, tunneling and transport properties for
a system with finite width  differ significantly from those of
a  semi-infinite model because of a gap in the energy dispersion.

\medskip

In the limiting semi-infinite geometry for a 2DTI, we obtain the
following exact 1D effective edge model along the
$y$-direction,\,\cite{Qi}

\begin{equation}
\mc{H}^{\rm eem}_{1D}=\mc{A}k_y\,\tensor{\sigma}_z\ ,
\end{equation}
whose  corresponding  dispersion relations are
 $\varepsilon_{1D}^{\rm eem}=\pm \mc{A} k_y$ with corresponding
  wave functions are the eigenfunctions of $\sigma_z$,
  i.e., $\Psi_1^{T}=\{ 0,\,1\}$ and $\Psi_2^{T}=\{ 1,\,0 \}$,
  which yield a transmission amplitude $\mc{T}_{\rm eem}=1$
  through a barrier of any height analogous to the Klein paradox
  for  head-on collisions in graphene. In contrast to the 2D model in
   Eq.\,\eqref{H2Dmain}, we find that the electron transmission
   for edge states with a finite width is substantially suppressed.

\medskip

Taking the limit $L_y\to \infty$, we can also consider bulk
states which are located far away from either edge.
In this case, we make the substitution $k_y = -i \partial_y$.
By retaining terms up to  order  ${\cal O}(k^3)$, calculation
 leads to a new energy dispersion for $\mc{A} \gg \mc{B}k$

\begin{equation}
\label{dbulk}
\varepsilon^{\rm bulk}_{2D}=-\mc{D}k^2 \pm \sqrt{\mc{M}^2+(\mc{A}^2-2\mc{M}\mc{B})k^2} \ .
\end{equation}
The Hamiltonian  in Eq.\,\eqref{H2Dmain2} assumes the
 the simple form  $\pm \mc{M}\,\tensor{\sigma}_z$ at ${\bf k}=0$,
where the gap parameter $\mc{M}$ strongly depends on the thickness
$d$ of the quantum well   and can be arbitrarily small or even set equal to
zero. The energy dispersion relations in Eq.\,\eqref{dbulk} formally
reduce to those of gapped graphene or graphene irradiated with
 circularly polarized light. The tunneling problems for this case were
 addressed  in Refs.\, [\onlinecite{Barb_review}] and
 [\onlinecite{mine}]. The most significant effect of radiation
 on  electron tunneling is the breaking of chiral
symmetry on the order of ${\cal O}(\mc{M}^2)$. Consequently,
significantly different behavior in the electron transmission at
small incident angles ($\vert k_y\vert \ll\vert k_x\vert$) is
expected compared to  infinite graphene in the absence of light illumination\,\cite{Katsnelson}.

\medskip

As derived from Ref. \cite{Finite_Size}, the decay rates
$\xi_{1,2}$ introduced in Eq.\,\eqref{mast} are given by

\begin{gather}
\label{la}
\xi_{1,2}=\frac{\mc{A}\sqrt{1-(\mc{D}/\mc{B})^2}}{2(\mc{B}+\mc{D})} \pm \frac{\sqrt{\mc{A}^2(\mc{B}-\mc{D})-4\mc{B}\mc{M}(\mc{B}+\mc{D})}}{2 \mc{B}\sqrt{\mc{B}+\mc{D}}} \ .
\end{gather}
The above equation, in conjunction with Eq.\,\eqref{mast},
leads to the following energy dispersion relations

\begin{equation}
\varepsilon_\beta + \frac{\mc{M}\mc{D}}{\mc{B}} = \beta \sqrt{\Delta_z^2+\left[{\frac{\mc{A}}{\mc{B}^2}
(\mc{B}^2-\mc{D}^2)\,k_x}\right]^2} + \mc{O}(k_x^2) \ ,
\end{equation}
where $\Delta_z$ is defined by \cite{Finite_Size}

\begin{equation}
\Delta_z=\frac{4 \mc{A}\mc{M}(\mc{B}^2-\mc{D}^2)}{\mc{B}^3[\mc{A}^2\mc{B}-4 \mc{M}(\mc{B}^2-\mc{D}^2)]} \ .
\end{equation}
Additionally, the transposed wave function associated with
$\xi_{1,2}$ under the limit of $\xi_{1,2}L_y\gg 1$ is

\begin{gather}
\Psi_\beta^T(y)=\frac{\texttt{e}^{i k_x x}}{\mc{N}} \{ f_{\rm e} - \beta s_k f_{\rm o},\, \frac{(\mc{B}+\mc{D})\,(\xi_1+\xi_2)}{\mc{A}}( f_{\rm o} + \beta s_k f_{\rm e}) \} \ ,
\end{gather}
with the notation $\beta=\pm 1$, $s_k=\pm 1$,  $\mc{N}$ is the
normalization factor, and we express the wave function components
in terms of odd ($f_{\rm o}$ with $-$) and even ($f_{\rm e}$ with $+$)
functions, defined by\,\cite{Finite_Size}

\begin{gather}
f_{{\rm e}({\rm o})}(y)=\frac{\texttt{e}^{\xi_1 y} \pm \texttt{e}^{-\xi_1 y}}{\texttt{e}^{\xi_1 L_y/2} \pm \texttt{e}^{-\xi_1 L_y/2}}
-\frac{\texttt{e}^{\xi_2 y} \pm \texttt{e}^{-\xi_2 y}}{\texttt{e}^{\xi_2 L_y/2} \pm \texttt{e}^{-\xi_2 L_y/2}} \ .
\end{gather}
\medskip

We now turn to a comparison of the electron tunneling in 2DTI
with graphene zigzag nanoribbon (ZNR) quasi-1D edge states.
In Ref.\, [\onlinecite{Fertig1}], the authors calculated
the ZNR wave functions as

\begin{equation}
\Psi^\beta_{\rm ZNR}(k_x,\,y)=
\left({ \begin{array}{c}   \beta \texttt{e}^{i \pi/2}
\sinh[\lambda_{y,n} (L_y/2+y)] \\  \hspace{0.37in}
\sinh[\lambda_{y,n}(L_y/2-y) ] \end{array} }
\right)  \ ,
\label{znrl}
\end{equation}
where $\lambda_{y,n}$ is related to the edge
distribution of electron wave functions. The boundary conditions
are such that each of the wave function components vanishes
 at one of the two ribbon edges (zigzag configuration). This
 leads to the following relation

\begin{equation}
\label{lznr}
\frac{k_x-\lambda_{y,n}}{k_x+\lambda_{y,n}}
=\texttt{e}^{-2 \lambda_{y,n}L_y} \ .
\end{equation}
Equation\ \eqref{lznr} demonstrates that the real solutions for
$\lambda_{y,n}$ exist only if both the conditions
$k_x>\lambda_{y,n}$ and $L_y\lambda_{y,n} >1$ are satisfied.
We will assume the ribbon is sufficiently wide so that
  $k_xL_y > 1$ for $k_x$ in the range of interest.
For real $\lambda_{y,n}$, the edge states decay  exponentially
  with decay length $1/\lambda_{y,n}$. This situation is similar
  to the case of 2DTI discussed above in this section. However,
  the traverse wave function distribution is drastically different
  from a graphene armchair nanoribbon (ANR) which has a plane-wave
  type wave function $\Psi_{\rm ANR}(y) \sim \texttt{e}^{i k_y y}$.
We also note from Eq.\,\eqref{lznr} that for large $k_x$ one gets $\lambda_{y,n}\approx 0$, i.e., fast moving electrons with large
longitudinal momenta have a negligible decay rate. Therefore, the
wave function in this case extends far away from the ribbon edges.
For chosen ribbon width, there exists a maximum value for the decay
rate.

\medskip

\begin{figure}[ht!]
\centering
\includegraphics[width=0.85\textwidth]{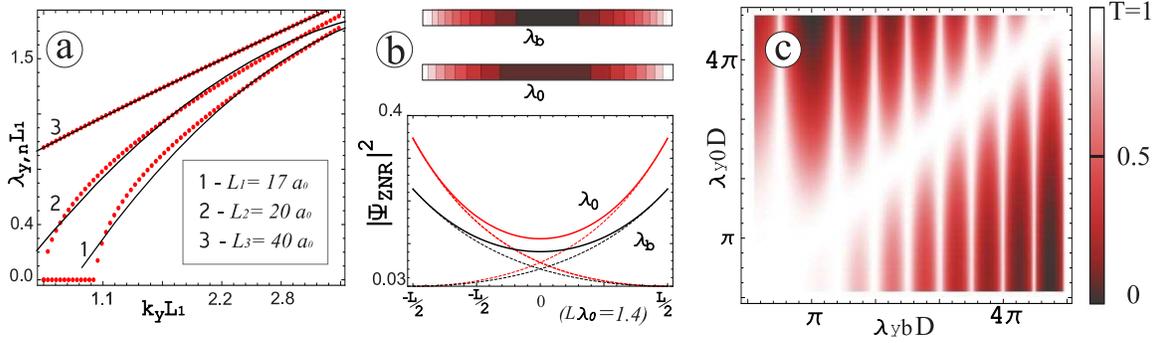}
\caption{(Color online) Zigzag nanoribbon wave function and tunneling
conductance. Panel $(a)$ demonstrates transverse decay rates $\lambda_{y,n}$
 depending on the longitudinal momentum $k_x$ for various nanoribbon widths.
 Panel $(b)$ shows the $y$-dependence of the edge state wave functions
  inside the barrier region ($\lambda_b$) and outside the barrier
  region ($\lambda_0$), as well as  contour plots of wave functions.
  Panel $(c)$ presents  density plot for the dependence of two-terminal
  conductance $g(\varepsilon)$ on $\lambda_0$ and $\lambda_b$.}
\label{znr}
\end{figure}

The energy dispersion for ZNR corresponding to Eq.\,eqref{znrl}
is given by $\varepsilon^{\rm ZNR}_{\beta}
= \beta \sqrt{k_x^2-\lambda_{y,n}^2}$. We confine our
attention to low potential barriers, such that $\lambda_{y,n}$
is real within a barrier region. Unimpeded tunneling in ZNR was
investigated in Ref.\, [\onlinecite{our}]. Additionally, we calculated
the two-terminal tunneling conductance by making use of

\begin{equation}
\label{ballis}
g(\varepsilon)=\frac{\mc{G}(\varepsilon)}{\mc{G}_0} = \frac{1}{L_y} \int\limits_{-L_y/2}^{L_y/2}\mc{T}(\varepsilon,\,y)\,d y \ ,
\end{equation}
showing how the ribbon conductivity is modified by a barrier region.

\medskip

From Fig.\,\ref{znr}(a), we find that the wave-function transverse
decay length $\lambda_{y,n}^{-1}$ decreases with increasing width
of a zigzag nanoribbon. Figure\ \ref{znr}(b) show us that the
minima of edge-state wave functions in the transverse direction
are at the center ($y=0$) for both inside and outside barrier
regions. Furthermore, the wave functions are symmetric with
respect to $y=0$. Our numerical results based on Eq.\,\eqref{ballis}
are presented in Fig.\,\ref{znr}(c). The diagonal line in
Fig.\,\ref{znr}(c) reflects the fact that $\mc{T}(\varepsilon,\,y)\equiv 1$
[or $g(\varepsilon)\equiv 1$] for the same decay rates $\lambda_0=\lambda_b$.
The periodic resonant peaks with respect to $\lambda_{b}$ can be
clearly seen in Fig.\,\ref{znr}(c) similar to the prediction by
Eq.\,\eqref{deltag}.

\section{Concluding remarks}
\label{conclu}

In summary, we have analytically obtained the energy dispersion
relations as well as the wave functions of electron dressed
states in TIs  irradiated by circularly polarized
light. A number of helical systems, such as graphene, nanoribbons
and topological insulators, are compared . Similar to graphene,
the electron-photon coupling in TI leads to a energy gap in the
electron energy dispersion relations and  eigenstates
with the broken chirality symmetry. The tunneling over a square
potential barrier is modified significantly if there exists a
gap in the  energy dispersion of electrons. The combination
of Schr\"odinger-like (massive) and Dirac-like (massless)
electrons tunneling below and above a critical energy gives
rise to very novel properties. We have further found that some
lower energy subbands become nearly dispersiveless with increasing
light intensity, which results in  unusual electronic properties
non-e in graphen

\medskip

As shown in Ref.\, [\onlinecite{Kibis}] for graphene, laser power
$10^2$\,W may produce an energy gap $\Delta \sim 10-100$\,meV
required for making the effect noticeable for THz light frequencies
at room temperature. Since the linear-term coefficient
(group velocities) of the 3DTI Hamiltonian has the same order
of magnitude as  graphene, we expect that an experimental verification
to be possible.

\medskip

Although the linear term in the energy dispersion of 3DTIs
gives spin-polarized Dirac cones with chirality of the
corresponding eigenstates, the appearance of a quadratic
term  introduces additional effects on electron tunneling.
Our calculations for  electron tunneling through a square
potential barrier indicate that  electrons may
be transmitted either like chiral particles as they do in
graphene or like conventional 2DEG electrons, depending on
the incident particle energy above or below a critical energy.

\medskip

We have also investigated  tunneling properties of edge
states for massive particles in 2DTIs, such as HgTe/CdTe
quantum wells, where an energy gap is introduced by the
finite width of samples\,\cite{Finite_Size}.
From the analysis of dressed edge-state wave function in
the presence of circularly polarized light, we have found
that particles can always freely propagate   along the ribbon
edges, which is  similar to   zigzag graphene nanoribbons
but different from semi-infinite 2DTIs where either  perfect
transmission or  complete reflection is obtained for
electron-electron  and electron-hole transitions, respectively.
Although the wave functions in both 2DTI and zigzag ribbons
are localized around the edges, the decay rate of edge state
wave functions of 2DTI does not depend on the longitudinal
wave number, in contrast to the decay rate in a zigzag ribbon.

\section*{ACKNOWLEDGEMENTS}

This research was supported by  contract \# FA 9453-11-01-0263
of AFRL. DH would like to thank the Air Force Office
of Scientific Research (AFOSR) for its support. The authors
also acknowledge considerable contribution and helpful
discussions with Liubov Zhemchuzhna.

\appendix

\section{Model of 3DTI  Surface  Irradiated with Circularly
Polarized Light}
\label{ap1}

Tthe Hamiltonian describing the surface states (at $z=0$) of a
3DTI to  order of ${\cal O}(k^2)$ is given by

\begin{equation}
\mc{H}^{\rm surf}_{3D} =\mc{D}k^2+\mc{A}\tensor{\sigma}\cdot{\bf k} = \left({  \begin{array}{cc}  \mc{D} k^2 & \mc{A}k_-  \\
\mc{A}k_+  &  \mc{D}k^2 \end{array} }\right) \ ,
\label{a1}
\end{equation}
where ${\bf k}=(k_x,\,k_y)$ is the un-plane surface wave vector
 and $k_{\pm}=k_x \pm i k_y$. The energy dispersion associated
 with this Hamiltonian is given by
 $\varepsilon^{\rm surf}_{3D}=\mc{D}k^2+\beta \mc{A}k$ with
 $\beta = \pm 1$.

\medskip

We  now turn to the case when the surface of the 3DTI is irradiated
 by circularly polarized light with  vector potential

\begin{equation}
\label{vpsurf}
\hat{\mathbf{A}}={\cal F}_0\left(\texttt{\bf e}_+\hat{a}+\texttt{\bf e}_-\hat{a}^{\dag}\right) \ ,
\end{equation}
where $\texttt{\bf e}_\pm=(\texttt{\bf e}_x \pm i \texttt{\bf e}_y)/\sqrt{2}$,
 $\texttt{\bf e}_x$ and $\texttt{\bf e}_y$ are   unit vectors in
 the $x$ and $y$ direction, respectively. Consequently, the in-plane
 components of the vector potential may be expressed as

\begin{gather}
\hat{A}_x=\frac{{\cal F}_0}{\sqrt{2}}(\hat{a}+\hat{a}^{\dag})\ ,
\hspace{0.25in} \hat{A}_y=i \frac{{\cal F}_0}
{\sqrt{2}}(\hat{a}-\hat{a}^{\dag}) \ .
\end{gather}
In order to include electron-photon coupling, we make the following
substitutions for electron wave vector

\begin{gather}
k_x \longrightarrow k_x +\frac{e\hat{A}_x}{\hbar} = k_x+\frac{e{\cal F}_0}{\sqrt{2}\hbar}(\hat{a}+\hat{a}^{\dag}) \ , \notag \\
k_y \longrightarrow k_y+\frac{e\hat{A}_y}{\hbar} = k_y+i \frac{e{\cal F}_0}{\sqrt{2}\hbar}(\hat{a}-\hat{a}^{\dag}) \ , \notag \\
k_+ \longrightarrow k_{+}+\frac{\sqrt{2}e{\cal F}_0}{\hbar}\,\hat{a}^{\dag},  \hspace{0.4in}  k_{-} \longrightarrow k_{-}+\frac{\sqrt{2}e{\cal F}_0}{\hbar}\,\hat{a} \ ,   \notag \\
k^2=k_+ k_- \longrightarrow k^2+\frac{\sqrt{2}e{\cal F}_0}{\hbar}\left(k_+\hat{a}+k_-\hat{a}^{\dag}\right) + \left(\frac{\sqrt{2}e{\cal F}_0}{\hbar}\right)^2\hat{a}^{\dag}\hat{a} \ .
\end{gather}
In our investigation, we consider high intensity  light with $N_0=\langle\hat{a}^\dag\hat{a}\rangle \gg 1$, and then,
$\hat{a}\hat{a}^{\dag}\sim\hat{a}^{\dag}\hat{a}$ due to
$\hat{a}\hat{a}^{\dag}= \hat{a}^{\dag}\hat{a} + 1$ for bosonic
operators. We adopt this simplification only for the second-order
terms $\sim \left(\sqrt{2}e{\cal F}_0/\hbar\right)^2$
but not for the principal ones containing $\hbar\omega_0$.
With the aid of these substitutions, the Dirac-like contribution
 to the Hamiltonian in Eq.\,\eqref{a1} becomes

\begin{gather}
\mc{H}_{\rm Dirac}= \mc{A}\tensor{\sigma}\cdot{\bf k} = \mc{A}\left(\tensor{\sigma}_-k_+ +\tensor{\sigma}_+k_-\right)  \notag \\
\longrightarrow \mc{A} \left(\tensor{\sigma}_-k_++\tensor{\sigma}_+k_-\right)+\frac{\sqrt{2}e{\cal F}_0}{\hbar}\,\mc{A}
\left(\tensor{\sigma}_-\hat{a}^{\dag}+\tensor{\sigma}_+\hat{a}\right)\ ,
\end{gather}
where $\tensor{\sigma}_{\pm}=(\tensor{\sigma}_x \pm
i \tensor{\sigma}_y)/2$. To describe a full electron-photon coupled
system, we also need to take into account  the photon energy
term $\hbar\omega_0\,\hat{a}^{\dag}\hat{a}$.  This yields

\begin{gather}
\hat{\mc{H}}=\left(\hbar\omega_0+4\mc{D}\zeta^2\right)
\hat{a}^{\dag}\hat{a}+\mc{D}k^2\tensor{\mathbb{I}}_{[2]}
+2\zeta \mc{D}\left(k_+\hat{a} + k_-\hat{a}^{\dag}\right)
\tensor{\mathbb{I}}_{[2]}   \notag \\
+ \mc{A}  \left(\tensor{\sigma}_+k_-+\tensor{\sigma}_-k_+\right)
+2\zeta\mc{A}\left(\tensor{\sigma}_+\hat{a}
+\tensor{\sigma}_-\hat{a}^{\dag}\right)\ ,
\label{hamil}
\end{gather}
where $\zeta=e{\cal F}_0/(\sqrt{2}\hbar)$. We may also rewrite
the Hamiltonian in Eq.\,\eqref{hamil} in matrix form as

\begin{gather}
\hat{\mc{H}} =\left(\hbar\omega_0+4\mc{D}\zeta^2\right)
\hat{a}^{\dag}
\hat{a}+\underline{\mathbf{1}}+\underline{\mathbf{2}}
+\underline{\mathbf{3}} \notag \\
\equiv\left(\hbar\omega_0+4\mc{D}\zeta^2\right)\hat{a}^{\dag}
\hat{a} +\left({  \begin{array}{cc}  \mc{D} k^2 & \mc{A}k_-  \\
\mc{A}k_+  &  \mc{D}k^2 \end{array} }\right) \notag \\
+2\zeta\mc{D} \left({  \begin{array}{cc} (k_-\hat{a}^{\dag}
+ k_+\hat{a})  & 0  \\
0  & (k_-\hat{a}^{\dag} +k_+\hat{a}) \end{array} }\right)
+2\zeta\mc{A} \left({  \begin{array}{cc}  0 & \hat{a}  \\
\hat{a}^{\dag}  &  0 \end{array} }\right)\ ,
\label{mainhamB}
\end{gather}
where $\underline{\mathbf{1}}\equiv\hat{\mc{H}}^{\rm surf}_{3D}$
denotes the  initial surface Hamiltonian with no electron-photon
interaction, $\underline{\mathbf{3}}$ gives  the principal
effect due to light coupled to electrons (the only non-zero term
at ${\bf k}=0$) and $\underline{\mathbf{2}}$ is the leading term
demonstrating the difference between dressed states in graphene
and 3DTI.
\medskip

We know from Eq.\,\eqref{mainhamB} that the Hamiltonian at
${\bf k}=0$ reduces to the exactly solvable Jayness-Cummings model,
after we neglect the field correction on the order of
${\cal O}(\zeta^2)$. We obtain

\begin{equation}
\label{JC}
\hat{\mc{H}}_{{\bf k}=0}=\hbar\omega_0\,\hat{a}^{\dag}
\hat{a} + 2\zeta \mc{A}\left(\tensor{\sigma}_+ \hat{a} +
\tensor{\sigma}_- \hat{a}^{\dag}\right) \ .
\end{equation}
Following the method adopted in   Refs.\ [\onlinecite{Kibis}] and
[\onlinecite{mine}], we expand the eigenfunctions of
Eq.\,\eqref{JC} as

\begin{gather}
\vert \Psi^0_{\uparrow, N_0} \rangle = \mu_{\uparrow, N_0} \vert \uparrow, N_0 \rangle +\nu_{\uparrow, N_0} \vert \downarrow, N_0+1 \rangle \ , \notag \\
\vert \Psi^0_{\downarrow, N_0} \rangle = \mu_{\downarrow, N_0} \vert \downarrow, N_0 \rangle - \nu_{\downarrow, N_0} \vert \uparrow, N_0-1 \rangle  \ .
\label{basisdef}
\end{gather}
By using the  properties

\begin{gather}
\hat{a}^{\dag}\,\vert \uparrow \downarrow, N_0 \rangle
= \sqrt{N_0+1}\,\vert \uparrow \downarrow, N_0+1 \rangle\ ,  \notag \\
\hat{a}\,\vert \uparrow \downarrow, N_0 \rangle=\sqrt{N_0}\,\vert \uparrow \downarrow, N_0-1 \rangle\ ,  \notag \\
\label{basis1}
\tensor{\sigma}_{\pm}\,\vert \downarrow\uparrow, N_0 \rangle
=  (1-\delta_{\uparrow, +})(1-\delta_{\downarrow, -})\,\vert
\uparrow \downarrow, N_0 \rangle\ ,
\end{gather}
we obtain the energy eigenvalues

\begin{gather}\
\frac{\varepsilon^0_{\pm}}{\hbar\omega_0}= N_0 \pm \frac{1}{2} \mp \frac{1}{2}\,\sqrt{1+\frac{\alpha^2}{N_0}\left(N_0 + \frac{1}{2} \pm \frac{1}{2}\right)} \notag \\
\simeq N_0 \pm \frac{1}{2} \mp
\left({\frac{1}{2} +\frac{1}{4}\,\alpha^2}\right)
= N_0\mp \frac{\alpha^2}{4}\ ,
\label{JCdisp}
\end{gather}
where $\alpha^2=2\zeta\mc{A}N_0/(\hbar\omega_0)$ with $N_0\gg 1$.
The energy gap at ${\bf k}=0$ has been calculated as
$\Delta^0\equiv\varepsilon^0_- - \varepsilon^0_+
\approx (\alpha^2/2)\,\hbar\omega_0$. We note that there is
no difference between graphene and the surface states of 3DTI
at ${\bf k}=0$, and therefore, the result in Refs.\, [\onlinecite{Kibis}]
 and   [\onlinecite{Gerry}] are relevant to each other.

\medskip

The  expansion coefficients in Eq.\,\eqref{basisdef} are calculated as

\begin{gather}
\mu_{\uparrow \downarrow, N_0} = \sqrt{\frac{1+\Sigma_{\uparrow \downarrow, N_0}}{2\,\Sigma_{\uparrow \downarrow, N_0}}}\ , \hspace{0.3in} \nu_{\uparrow
 \downarrow, N_0} = \sqrt{\frac{1-\Sigma_{\uparrow \downarrow, N_0}}{2\,\Sigma_{\uparrow \downarrow, N_0}}} \ ,
\end{gather}
where

\begin{equation}
\Sigma_{\uparrow \downarrow, N_0} = \sqrt{1-\frac{\alpha^2}{N_0} \left({N_0+\frac{1}{2} \pm \frac{1}{2}}\right)}  \ .
\end{equation}
In all further calculations in this Appendix, we assume $\alpha \ll 1 $
and $N_0 \gg 1$, corresponding to a larger number of lase photons
but weak light coupling to electrons as a perturbation to the electron
energy. This leads to the following approximate expressions

\begin{gather}
\mu_{\uparrow, N_0} \simeq \mu_{\downarrow, N_0}
\simeq \mu_{\uparrow, N_0+1}  \simeq \mu_{\downarrow, N_0+1}
\simeq 1 - \frac{\alpha^2}{4} \ , \\
\nu_{\uparrow, N_0} \simeq \nu_{\downarrow, N_0}
\simeq \nu_{\uparrow, N_0+1}  \simeq \nu_{\downarrow, N_0+1}
\simeq \frac{\alpha}{2}  \ .
\end{gather}
Consequently, we only need to keep one pair of index-free coefficients
$\{\mu, \nu \}$ such that $\mu=\cos(\Phi)$, $\nu=\sin(\Phi)$ with $\Phi=\tan^{-1}(\alpha/2)$.
Furthermore, from the first equity in Eq.\,\eqref{JCdisp}, it follows
 that the energy gap still depends on $N_0$ in general. However,
 the difference $\Delta^0_{N_0+1}-\Delta^0_{N_0}
 \sim \Delta^0/N_0$ is so small that we neglect it for $N_0 \gg 1$.

\medskip

Generalizing Eq.\,\eqref{basisdef}, we still expand the wave function $\Psi_{\mathbf{k}}$ over the eigenstates of the Hamiltonian
in Eq.\,\eqref{JC} for ${\bf k}\neq 0$, i.e.

\begin{gather}
\vert \Psi_{\mathbf{k}} \rangle =\sum\limits_{j=N_0}^{N_0+1} \sum\limits_{\uparrow,\downarrow} \Xi^{\bf k}_{\uparrow \downarrow, j} \vert \Psi^0_{\uparrow \downarrow, j} \rangle  \notag \\
=\Xi^{\bf k}_{\uparrow, N_0} \vert \Psi^0_{\uparrow, N_0} \rangle + \Xi^{\bf k}_{\downarrow, N_0} \vert \Psi^0_{\downarrow, N_0} \rangle +\Xi^{\bf k}_{\uparrow, N_0+1} \vert \Psi^0_{\uparrow, N_0+1} \rangle +\Xi^{\bf k}_{\downarrow, N_0+1} \vert \Psi^0_{\downarrow, N_0+1} \rangle  \ .
\label{a16}
\end{gather}
Using Eq.\,\eqref{a16} we project the full Hamiltonian in
Eq.\,\eqref{hamil} onto the representation
$\{\Xi^{\bf k}_{\uparrow, N_0},\Xi^{\bf k}_{\downarrow, N_0},
\Xi^{\bf k}_{\uparrow, N_0+1},\Xi^{\bf k}_{\downarrow, N_0+1}\}$.
This yields

\begin{equation}
\label{4ham}
\left({\begin{array}{cc|cc}
N_0\,\hbar \omega_0  + \mc{D} k^2- \Delta
& \mu^2 \mc{A} k_{-}  & \mu\nu \mc{A} k_{+} - 2\zeta \mc{D} k_+  & 0 \\
\mu^2\mc{A} k_{+} & N_0\,\hbar \omega_0 + \mc{D} k^2  + \Delta
& 0 & -2 \nu \mu \mc{A} k_{+} - 2\zeta \mc{D} k_+\\ \hline
\nu \mu \mc{A} k_{-} - 2\zeta \mc{D} k_- & 0
& (N_0+1) \,\hbar \omega_0 + \mc{D} k^2 - \Delta & \mu^2 \mc{A} k_{-} \\
0 & - 2\nu \mu \mc{A} k_{-} - 2\zeta \mc{D} k_- & \mu^2 \mc{A} k_{+} & (N_0+1)\,\hbar
\omega_0 + \mc{D} k^2  + \Delta  \end{array} }\right) \ ,
\end{equation}
where $\Delta\approx\Delta^0$ is the energy gap at ${\bf k}\neq 0$
and we have employed the relations for $N=N_0$ or $N_0+1$

\begin{gather}
\tensor{\sigma} \cdot {\bf k}\, \vert \Psi^0_{\uparrow N} \rangle
=\tensor{\sigma} \cdot {\bf k}\, \left\{{ \mu \vert \uparrow, N \rangle
+ \nu \vert \downarrow, N+1 \rangle }\right\} = \mu k_+ \vert
\downarrow, N \rangle + \nu k_- \vert \uparrow, N+1 \rangle  \ , \\
\tensor{\sigma} \cdot {\bf k}\,\vert \Psi^0_{\downarrow N} \rangle
=\tensor{\sigma} \cdot {\bf k}\, \left({ \mu \vert \downarrow, N \rangle
- \nu \vert \uparrow, N-1 \rangle }\right) = \mu k_- \vert \uparrow, N \rangle
- \nu k_+ \vert \downarrow, N-1 \rangle \ ,\\
\hat{a}\, \vert \Psi^0_{\uparrow \downarrow, N} \rangle = \sqrt{N}\,\vert \Psi_{\uparrow \downarrow, N-1} \rangle\ , \hspace{0.3in}
\hat{a}^{\dag}\,\vert \Psi^0_{\uparrow \downarrow, N} \rangle = \sqrt{N+1}\,\vert \Psi_{\uparrow \downarrow, N+1} \rangle  \ .
\end{gather}
		
\medskip

After calculating the eigenvalues for the Hamiltonian
in Eq.\,\eqref{4ham}, we  obtain closed form analytic expressions
with energy dispersion

\begin{gather}
\varepsilon_{\{N_0 \uparrow, N_0 \downarrow, N_0+1 \uparrow, N_0+1 \downarrow\}}(k,\,\Delta)=\left({N_0 +\frac{1}{2}}\right) \hbar \omega_0 + \mc{D} k^2 \pm \sqrt{\mathbb{C}_1(k,\,\Delta)\pm \sqrt{\mathbb{C}_2(k,\,\Delta)}} \ , \\
\mathbb{C}_1(k,\,\Delta)=\left({\hbar \omega_0 / 2}\right)^2 +\zeta \nu \mc{A} \mc{D} k^2 + \left[\zeta^2 \mc{D}^2 + \mc{A}^2(1 + 5/2 \nu^2 )\right] k^2  \ , \\
\mathbb{C}_2(k,\,\Delta)=\zeta^2 \mc{A}^2 \mc{D}^2 k^4 (1+ \nu^2) + 4 \nu (\zeta \nu \mc{D}+\mc{A}) \mc{A}^3 k^4 \notag \\
 - 3 \nu (2 \zeta \mc{D}+\nu \mc{A}) \mc{A} \hbar \omega_0 \Delta k^2 + 4 \hbar^2 \omega_0^2 (\Delta^2 + \mc{A}^2 k^2) \ .
\end{gather}
Taking the limits $\zeta \mc{D} \to 0$ and  $\nu k\to 0$,
we are left with two uncoupled energy subbands

\begin{gather}
\varepsilon_{\{N_0 \uparrow, N_0 \downarrow\}}(k,\,
\Delta) = N_0\,\hbar \omega_0 + \mc{D} k^2 \pm \sqrt{
\Delta^2+ \mc{A}^2 k^2}  \ , \\
\varepsilon_{\{N_0+1 \uparrow, N_0+1 \downarrow\}}(k,\,
\Delta) = (N_0+1)\,\hbar \omega_0 + \mc{D} k^2 \pm
\sqrt{\Delta^2+ \mc{A}^2 k^2} \ .
\end{gather}
Therefore, we conclude that the effect of electron-photon
interaction is quite similar to graphene as far as one photon
number $N_0$ is concerned. The main difference being that
the energy gap in the 3DTI is of  order  ${\cal O}(\zeta^2)$,
which may be neglected for low intensity light. Consequently,
the energy dispersion relation becomes

\begin{equation}
\varepsilon_{\beta}(k,\,\Delta_0) = N_0\,\hbar \omega_0 + \mc{D} k^2 + \beta \sqrt{\left[\Delta_0 + {\cal O}(\zeta^2)\right]^2
 + \left({\mc{A}k}\right)^2} \ ,
\end{equation}
where $\beta=\pm 1$ and $\Delta_0$ is the photon-induced energy
gap as in graphene.

\section{Dressed  3DTI Electron surface states}
\label{ap2}

It follows from Appendix\ \ref{ap1} that the effect due to
$\sim \hat{\bf A}^2$-terms would play a the role only if two
optical modes are considered. This effect is of   order
 ${\cal O}(\zeta^2)$ and may be neglected in our calculations.
When the surface of a 3DTI is irradiated with circularly polarized
light, the light will penetrate into the sample and
decay exponentially away from the surface, similar to the wave
function for a surface electronic state. Therefore, by including
this decay effect, the vector potential in Eq.\,\eqref{vpsurf}
is generalized as

\begin{equation}
\label{vpbulk}
\hat{{\bf A}}={\cal F}_0\left(\texttt{\bf e}_+\hat{a}+\texttt{\bf e}_-\hat{a}^{\dag}\right)\,\texttt{e}^{-\xi z}\ ,
\end{equation}
where $1/\xi$ is the decay length. By including the $z$ dependence,
the Hamiltonian of the electron system is found to be

\begin{gather}
\hat{\mc{H}}_{3D}({\bf k}_\bot,\,z)=
\hat{\mc{H}}^{(1)}_{3D}(z)+\hat{\mc{H}}^{(2)}_{3D}({\bf k}_\bot) \ ,
\label{tota}
\end{gather}
where

\begin{gather}
\hat{\mc{H}}^{(1)}_{3D}(z) =\left({\mc{C}-\mc{D}_z\partial_z^2}\right)\tensor{\mathbb{I}}_{[4]}+\left({ \begin{array}{cc}    (\mc{M}+\mc{B}_z\partial_z^2)\, \tensor{\sigma}_3 - i \mc{A}_{z}\partial_z\,\tensor{\sigma}_1  &  0 \\
 0  &  (\mc{M}+\mc{B}_z\partial_z^2)\,\tensor{\sigma}_3 + i \mc{A}_{z}\partial_z\,\tensor{\sigma}_1
\end{array} }\right) \ , \\
\hat{\mc{H}}^{(2)}_{3D}({\bf k}_\bot) = -\mc{D}_{\bot} k^2\tensor{\mathbb{I}}_{[4]} -  \mc{B}_{\bot} k^2\,\tensor{\sigma}_3 \otimes\tensor{\mathbb{I}}_{[2]}   +  \left({ \begin{array}{cc}                     0                &            \mc{A}_{\bot}k_-\,\tensor{\sigma}_1 \\
\mc{A}_{\bot}k_+\, \tensor{\sigma}_1                &            0
\end{array} }\right) \ .
\label{b14}
\end{gather}
Here, ${\bf k}_\bot=(k_x,\,k_y)$,
$\mc{C}$, $\mc{M}$, $\mc{A}_z$ , $\mc{B}_z$,
$\mc{D}_z$ are parameters in the Kane ${\bf k}\cdot{\bf p}$
model for bulk states, and $\mc{A}_\bot$ , $\mc{B}_\bot$,
$\mc{D}_\bot$ are parameters for surface states. We note that
the parameter $\mc{C}$ may be eliminated  by simply shifting the
band edges. Since the electron-photon interaction occurs in 
the region close to the surface, we only need to consider
$\hat{\mc{H}}^{(2)}_{3D}({\bf k}_{\bot})$ in Eq.\,\eqref{tota}, 
which contains the coupling with the incident circularly polarized 
light.

\medskip

When ${\bf k}=\{{\bf k}_\bot,\,k_z\}=0$, the Hamiltonian in 
Eq.\,\eqref{b14}  plus the single-photon energy as well as the 
light-electron coupling   together give 

\begin{equation}
\hat{\mc{H}}_{{\bf k}=0}=\hbar \omega_0\,\hat{a}^{\dag}\hat{a}+2\zeta\mc{A}_\bot\left(\begin{array}{cc}
 0 & \scriptsize{\begin{array}{cc}
0 & \hat{a} \\
\hat{a} & 0  \\
\end{array}}  \\
 \scriptsize{\begin{array}{cc}
0 & \hat{a}^{\dag} \\
\hat{a}^{\dag} & 0 \\
\end{array}} & 0 \end{array} \right)  \ .
\label{matrx}
\end{equation}
By introducing a small dimensionless parameter 
$\alpha^\prime=2\zeta\mc{A}_\bot / \mc{M} \ll 1$, and the 
two matrices

\begin{equation}
\tensor{\Gamma}_+=\tensor{\sigma}_{+} \otimes\tensor{\sigma}_1 = \left(\begin{array}{cc}
0 & \scriptsize{\begin{array}{cc}
0 & 1 \\
1 & 0 \\
\end{array} } \\
\scriptsize{ \begin{array}{cc}
0 & 0 \\
0 & 0 \\
\end{array}} & 0 \end{array} \right) \ ,
\hspace{0.3in}
\hat{\Gamma}_-=\tensor{\sigma}_{-} \otimes \tensor{\sigma}_1 = \left(\begin{array}{cc}
 0 & \scriptsize{\begin{array}{cc}
0 & 0 \\
0 & 0 \\
\end{array} } \\
\scriptsize{ \begin{array}{cc}
0 & 1 \\
1 & 0 \\
\end{array}} & 0 \end{array} \right)\ ,
\end{equation}
the Hamiltonian in Eq.\,\eqref{matrx} may be rewritten  
  compactly as    

\begin{equation}
\hat{\mc{H}}_{{\bf k}=0}=\hbar \omega_0\,\hat{a}^{\dag}\hat{a} + \alpha^\prime(\tensor{\Gamma}_-\hat{a}^{\dag}+\tensor{\Gamma}_+\hat{a}) \ .
\end{equation}
In  analogy with Eq.\,\eqref{basisdef}, we construct a basis
 set containing four states, i.e., $\vert++\,\rangle 
 =\vert 1,0,0,0 \rangle$, $\vert +-\rangle =\vert 0,1,0,0 \rangle$, 
 $\vert - +\rangle =\vert 0,0,1,0 \rangle$, 
 $\vert - -\rangle =\vert 0,0,0,1 \rangle$. For this basis set, it is 
 a simple matter to show  the following properties

\begin{gather}
\hat{a}^{\dag}\,\vert N,  \pm \pm \rangle  
= \sqrt{N+1}\,\vert N+1,  \pm \pm \rangle, \notag \\
\hat{a}\,\vert N,  \pm \pm \rangle = \sqrt{N}
\,\vert N-1,  \pm \pm \rangle, \notag \\
\tensor{\Gamma}_+\,\vert N, + \pm \rangle =0 \ ,  \hspace{0.4in} \notag \tensor{\Gamma}_+ \vert N, + \pm \rangle =0  \ , \\
\tensor{\Gamma}_+\,\vert N, - \pm \rangle 
=\vert N, + \mp \rangle \ , \hspace{0.4in} 
\tensor{\Gamma}_- \vert N, + \mp \rangle =\vert N, - \pm \rangle  \ .
\end{gather}
We may also expand the dressed electronic states at ${\bf k}=0$ 
over this basis set leading to

\begin{gather}
\vert \Psi^0_{++,N} \rangle = \mu_{N,+} \vert ++, N \rangle 
+ \nu_{N,+} \vert +-, N \rangle + \nu_{N+1,-} \vert -+, N+1 \rangle 
+ \nu_{N+1,-} \vert --, N+1 \rangle  \ ,\notag \\
\vert \Psi^0_{+-,N} \rangle = \nu_{N,+} \vert ++, N \rangle 
+ \mu_{N,+} \vert +-, N \rangle + \nu_{N+1,-} \vert -+, N+1 \rangle 
+ \nu_{N+1,-} \vert --, N+1 \rangle \ , \notag \\
\vert \Psi^0_{-+,N} \rangle = \nu_{N,+} \vert ++, N \rangle 
+ \nu_{N,+} \vert +-, N \rangle + \mu_{N+1,-} \vert -+, N+1 \rangle
 + \nu_{N+1,-} \vert --, N+1 \rangle \ ,\notag \\
\vert \Psi^0_{--,N} \rangle = \nu_{N,+} \vert ++, N \rangle 
+ \nu_{N,+} \vert +-, N \rangle + \nu_{N+1,-} \vert -+, N+1 \rangle 
+ \mu_{N+1,-}
\vert --, N+1 \rangle \ . \notag
\end{gather}
For $N \gg 1$ and $\alpha^\prime\ll 1$,   calculation shows that

\begin{gather}
\mu_{N, +} \simeq \mu_{N,-} \simeq \mu_{N+1, +}  
\simeq \mu_{N+1, -} \simeq 1 - \left(\frac{\alpha^{\prime}}{2}\right)^2 
\ , \\
\nu_{N} \simeq \nu_{N, -} \simeq \nu_{N+1. +}  \simeq \nu_{N+1, -} \simeq \frac{\alpha^\prime}{2} \ .
\end{gather}

\medskip

Introducing the pair of operators

\begin{gather}
\tensor{\Gamma}_{12}=\left(\begin{array}{cc}
\scriptsize{\begin{array}{cc}
0 & 1 \\
0 & 0 \\
\end{array} }  & 0 \\
0 & 0
\end{array} \right) \ , \hspace{0.3 in} 
\tensor{\Gamma}_{34}=\left(\begin{array}{cc}
0  & 0 \\
0 &
\scriptsize{ \begin{array}{cc}
0 & 1 \\
0 & 0 \\
\end{array}} \end{array} \right)\ ,
\end{gather}
we obtain

\begin{gather}
\tensor{\Gamma}_{12}\,\vert N, +- \rangle = \vert N, ++ \rangle \ , \hspace{0.3in} \tensor{\Gamma}_{34}\,\vert N, ++ \rangle = \vert N, +- \rangle \ ,  \\
\tensor{\Gamma}_{34}\,\vert N, -+ \rangle = \vert N, -- \rangle \ , \hspace{0.3in} \tensor{\Gamma}_{12}\,\vert N, -- \rangle = \vert N, -+ \rangle \ .
\end{gather}
For the Hamiltonian in Eq.\,\eqref{mainhamB} within the 
$2\times 2$ subspace, as a special case, we  introduce

\begin{equation}
\tensor{\sigma}_+=\tensor{\Gamma}_{12} \ , \hspace{0.3in} \tensor{\sigma}_-=\tensor{\Gamma}_{34}  \ .
\end{equation}
Finally, we can rewrite the full Hamiltonian in Eq.\,\eqref{tota} 
for ${\bf k}\neq 0$, using the basis set $\vert N_0, \pm \pm \rangle$, as

\begin{gather}
\label{3Dmaina}
\hat{\mc{H}}_{3D}({\bf k}_\bot,\,z)=N_0\,\hbar \omega_0+   \notag \\
\scriptsize{  \left({ \begin{array}{cc|cc}     \mc{C_+}-\mc{D}_{z,-}\partial_z^2 -\mc{D}_{\bot,-} k^2 - \Delta &                             -i \mc{A}_{z}\partial_z &                                       0 &                                                \mc{A}_{\bot}k_- \\
-i \mc{A}_{z}\partial_z &                  \mc{C_-}-\mc{D}_{z,+}\partial_z^2 -\mc{D}_{\bot,+} k^2   - \Delta  &                         \mc{A}_{\bot}k_- &                                                        0 \\  \hline
\mc{A}_{\bot}k_+ &                                      \mc{C_+}-\mc{D}_{z,-}\partial_z^2 -\mc{D}_{\bot,-} k^2  + \Delta &                      i \mc{A}_{z}\partial_z \\
\mc{A}_{\bot}k_+ &                                              0 &                                       i \mc{A}_{z}\partial_z                      &              \mc{C_-}-\mc{D}_{z,+}\partial_z^2 + \mc{D}_{\bot,+} k^2  + \Delta
\end{array} }\right)   } \ ,
\label{3deq}
\end{gather}
where the $\pm$ signs correspond to opposite pseudo-spins in 
the basis set. Making use of Eq.\,\eqref{3deq}, we obtain the secular 
equation for the energy dispersion relations when $\Delta=0$, that is,

\begin{equation}
\varepsilon^2-\mc{A}_{\bot}^2 k^2+ \mathbb{N}_1 \mathbb{N}_1+\mc{A}_z^2 \xi^2 + \mc{D}_- \mathbb{N}_1 \xi^2+\mc{D}_+ \mathbb{N}_2 \xi^2+\mc{D}_- + \mc{D}_+ \mathbb{N}_2 \xi^2 + \mc{D}_+ \mc{D}_- \xi^4 -(\mathbb{N}_1+\mathbb{N}_2 + 2\mc{D} \xi^2)\,\varepsilon  = 0 \ ,
\label{dispe}
\end{equation}
where we have introduced the notations

\begin{gather}
\mathbb{N}_{1,2}(k)=\mc{C} \pm \mc{M} + (\mc{D}_{\bot} \mp \mc{B}_{\bot})\,k^2 \ , \\
\mc{D}_{\pm} = \mc{D}_z \pm \mc{B}_z  \ .
\end{gather}

\bibliographystyle{unsrt}
\bibliography{TItunBiblio}

\begin{thebibliography}{10}

\bibitem{HasanTI}
M.~Z. Hasan and C.~L. Kane.
\newblock \textit{Colloquium} : Topological insulators.
\newblock {\em Rev. Mod. Phys.}, 82:3045--3067, Nov 2010.

\bibitem{Qi}
X.-L. Qi and S.-C. Zhang.
\newblock Topological insulators and superconductors.
\newblock {\em Rev. Mod. Phys.}, 83:1057--1110, Oct 2011.

\bibitem{Ber}
B.A. Bernevig, T.~A. Hughes, and S.-C. Zhang.
\newblock Quantum spin hall effect and topological phase transition in hgte
  quantum wells.
\newblock {\em Science}, 314:1751--1761, Dec. 2006.

\bibitem{Novoselov-main}
Geim A. K. Morozov S. V. Jiang D. Katsnelson M. I. Grigorieva I. V. Dubonos S.
  V. Firsov A.~A. Novoselov, K.~S.
\newblock Two-dimensional gas of massless dirac fermions in graphene.
\newblock {\em Nature}, 438(197), 2005.

\bibitem{geim}
A.~K. Geim.
\newblock Graphene: Status and prospects.
\newblock {\em Science}, 324(5934), 2009.

\bibitem{main_model}
W.-Y. Shan, H.-Z Lu, and S.-Q. Shen.
\newblock Effective continuous model for surface states and thin films of
  three-dimensional topological insulators.
\newblock {\em New Journal of Physics}, 12(4):043048, 2010.

\bibitem{Finite_Size}
B.~Zhou, H.-Z. Lu, R.-L. Chu, S.-Q. Shen, and Q.~Niu.
\newblock Finite size effects on helical edge states in a quantum spin-hall
  system.
\newblock {\em Phys. Rev. Lett.}, 101:246807, Dec 2008.

\bibitem{aoki_oki}
T.~Oka and H.~Aoki.
\newblock Photovoltaic hall effect in graphene.
\newblock {\em Phys. Rev. B}, 79(8):081406, 2009.

\bibitem{Kibis}
O.~V. Kibis.
\newblock Metal-insulator transition in graphene induced by circularly
  polarized photons.
\newblock {\em Phys. Rev. B}, 81(16):165433, 2010.

\bibitem{Kibis2}
O.~V. Kibis, O.~Kyriienko, and I.~A. Shelykh.
\newblock Band gap in graphene induced by vacuum fluctuations.
\newblock {\em Phys. Rev. B}, 84:195413, Nov 2011.

\bibitem{K3}
O.~V. Kibis.
\newblock Dissipationless electron transport in photon-dressed nanostructures.
\newblock {\em Phys. Rev. Lett.}, 107:106802, Aug 2011.

\bibitem{Katsnelson}
M.~I. Katsnelson, K.~S. Novoselov, and A.~K. Geim.
\newblock Chiral tunnelling and the klein paradox in graphene.
\newblock {\em Nat Phys}, 2:620--625, 2006.

\bibitem{mine}
Iurov A, G.~Gumbs, O.~Roslyak, and D.~H. Huang.
\newblock Anomalous photon-assisted tunneling in graphene.
\newblock {\em Journal of Physics: Condensed Matter}, 24(1):015303, 2012.

\bibitem{magntr}
S.~Mondal, D.~Sen, K.~Sengupta, and R.~Shankar.
\newblock Magnetotransport of dirac fermions on the surface of a topological
  insulator.
\newblock {\em Phys. Rev. B}, 82:045120, Jul 2010.

\bibitem{DSTransp}
D.~Culcer, E.~H. Hwang, T.~D. Stanescu, and S.~Das~Sarma.
\newblock Two-dimensional surface charge transport in topological insulators.
\newblock {\em Phys. Rev. B}, 82:155457, Oct 2010.

\bibitem{castroneto}
A.~H. Castro~Neto, F.~Guinea, N.~M.~R. Peres, K.~S. Novoselov, and A.~K. Geim.
\newblock The electronic properties of graphene.
\newblock {\em Rev. Mod. Phys.}, 81(1):109--162, 2009.

\bibitem{andreevr}
Y.~Tanaka, T.~Yokoyama, and N.~Nagaosa.
\newblock Manipulation of the majorana fermion, andreev reflection, and
  josephson current on topological insulators.
\newblock {\em Phys. Rev. Lett.}, 103:107002, Sep 2009.

\bibitem{Barb_review}
M.~Barbier, P.~, Vasilopoulos, and F.~M. Peeters.
\newblock Single-layer and bilayer graphene superlattices: collimation,
  addition dirac points and dirac lines.
\newblock {\em Philosophical Transactions of Royal Society A}, 386:5499--5524,
  2010.

\bibitem{Tw}
J.~Tworzyd\l{}o, B.~Trauzettel, M.~Titov, A.~Rycerz, and C.~W.~J. Beenakker.
\newblock Sub-poissonian shot noise in graphene.
\newblock {\em Phys. Rev. Lett.}, 96:246802, Jun 2006.

\bibitem{Flugge}
Siegfried Fl\"{u}gge.
\newblock Practical quantum mechanics.
\newblock {\em Springer Study Edition.}, 2009.

\bibitem{Konig_exp}
M.~K\"{o}nig, H.~Buhmann, L.~W. Molenkamp, T.~Hughes, C.-X. Liu, X.-L. Qi, and
  S.-C. Zhang.
\newblock The quantum spin hall effect: Theory and experiment.
\newblock {\em Journal of the Physical Society of Japan}, 77(3):031007, 2008.

\bibitem{Fertig1}
L.~Brey and H.~A. Fertig.
\newblock Electronic states of graphene nanoribbons studied with the dirac
  equation.
\newblock {\em Phys. Rev. B}, 73:235411, Jun 2006.

\bibitem{our}
O~Roslyak, A~Iurov, Godfrey Gumbs, and Danhong Huang.
\newblock Unimpeded tunneling in graphene nanoribbons.
\newblock {\em Journal of Physics: Condensed Matter}, 22(16):165301, 2010.

\bibitem{Gerry}
Christopher Gerry and Peter Knight.
\newblock Introductory quantum optics.
\newblock {\em Cambridge University Press}, 2005.

\end{thebibliography}

\end{document}